**Title:** Dirac excited state quenching in graphene

**Authors:** Jacky C. Wan[1,†], Trevor B. Arp[1,†], Nathaniel M. Gabor[1*]

**Affiliations:**
[1]Laboratory of Quantum Materials Optoelectronics, Department of Physics and Astronomy
University of California, Riverside; Riverside CA 92521, USA.

* Corresponding author. email: nathaniel.gabor@ucr.edu
† These authors contributed equally to this work

**Abstract:** Hot, dense phases of Dirac fermions - predicted to resemble relativistic plasma - are uniquely accessible through photoexcitation of pristine, charge neutral graphene. We demonstrate a sensitive temperature probe of the photoexcited Dirac state, called interlayer optoelectronic thermometry, which measures out-of-plane transport of hot carriers in high-mobility, neutral graphene encapsulated within graphene-hBN-graphene heterostructures. At a critical intermediate sample temperature $T = 50$ K, the electronic temperature $T_e$ is quenched, exhibiting an intrinsic cooling rate that exceeds $10^{14}$ Kelvin/s within the first picosecond after photoexcitation. Quenching is further enhanced by applying in-plane voltages within the stack-engineered heterostructure. Extreme sensitivity of $T_e$ to sample temperature and applied voltages reveals anomalously efficient hot-carrier quenching, which we identify as an essential feature of the strongly interacting hot Dirac excited state.

**Main Text:**
Graphene at charge neutrality is predicted to host a hot, dense electron-hole (e-h) excited state, which theoretically exists over a large range of electronic temperatures[1-3]. Several experimental observations at low electronic temperatures - in which electrons and holes are in near-equilibrium with the crystal lattice - have attributed unusual device response to this apparent Dirac plasma, through which energy may be transported with remarkable efficiency[1,4-7]. The violation of the conventional Wiedemann-Franz law[8], fluid-like behavior consistent with viscous electron flow[5,9,10], interparticle scattering rates limited by relativistic hydrodynamics to the shortest possible timescale for energy relaxation[1,2,4-7], enhanced thermal diffusivity at room temperature[11], and giant parabolic magneto-resistivity[12] have all been observed within this unconventional transport regime.

In this work, we report extremely efficient, anomalous quenching of the Dirac photo-excited state. Quenching is found to be enhanced at an intermediate sample temperature, despite the orders-of-magnitude disparity between the initial hot electronic temperature $T_e > 2000$ K and the sample $T = 50$ K. Importantly, this enhanced quenching process - which is highly sensitive to sample temperature - is unexpected based on known energy relaxation pathways: On one hand, sensitivity to sample temperature generally indicates relaxation through strong electron-vibrational coupling, yet these processes in graphene occur at time scales too long to quench the system efficiently[13-15]. On the other hand, a remarkable order-of-magnitude increase in the electronic thermal conductivity $\kappa_e$ at sample temperatures $T = 50 - 80$ K [8] has been observed in the Dirac fluid regime. At higher sample temperatures, and thus higher $T_e$, enhanced thermal conductivity was *not* observed, and is thus not expected to efficiently cool hot photoexcited carriers. Rapid quenching in high-mobility graphene heterostructures indicates an emergent cooling regime within the Dirac excited state and raises important questions about hydrodynamic energy transport in graphene.

Our observations are based on interlayer optoelectronic thermometry, which we developed to directly probe the Dirac e-h excited state in the uniform two-dimensional plane of charge neutral, high-mobility graphene. This excitation scheme allows us to generate a photocurrent *without* the need for a finite charge density or an in-plane p-n junction, which would influence the energy relaxation pathways[16]. After photoexcitation by short optical pulses, electrons and holes form a thermal distribution through graphene's rapid thermalization processes, prior to coming to equilibrium with the crystal lattice[17-20]. By introducing an energy barrier for *out-of-plane* charge transport - imparted by a graphene/hBN interface - hot charge carriers at the top of the thermal distribution are filtered out, giving rise to strong interlayer photo-thermionic current [21–24]. This interlayer photocurrent, which is exponentially sensitive to the number of charge carriers in the high-energy tail of the distribution, provides a high-resolution probe of the excited state temperature $T_e$ within the spatially uniform graphene layer (see supplemental methods).

The devices studied here consist of two graphene sheets, labeled $G_T$ and $G_B$ (top and bottom respectively), separated by a thin hexagonal boron nitride (hBN) layer (Fig. 1a). The hBN thickness ($L$ = 8-10 nm) is specifically chosen to fully suppress graphene-to-graphene tunneling[24] and photon assisted-tunneling current[25] at low voltages. The constituent layers, including the hBN encapsulants (not shown schematically), were mechanically laminated via an inverted dry transfer method using polymer stamps (see supplemental methods). These devices operate by applying a voltage $V_B$ to the bottom layer $G_B$, while probing the photocurrent $I_{PC}$ in the top layer $G_T$. For the data shown in Figs. 1 and 2 the top layer voltage $V_T$ = 0 V so that $V_B$ controls the voltage drop between graphene layers.

We generate electron-hole excited states using ultra-short, infrared optical gating pulses. A femtosecond (180 fs) optical parametric oscillator with wavelength of 1200 nm photoexcites the G/hBN/G devices with an optical pulse every 13 ns (see supplemental methods). Each pulse triggers a short burst of charge carriers that transit the hBN layer. The transit of these charge carriers is measured as an average interlayer photocurrent $I_{PC}$. Photocurrent imaging (Fig. 1b) shows a strong photoresponse only at the overlapping region; this is consistent with a purely interlayer photocurrent. This photocurrent (measured along the dashed line in Fig. 1b) increases super-linearly with incident laser power $P$ (Fig. 1c), and from a plot of $I_{PC}$ vs. $P$ in Fig. 1d, we find that the data is well described by $I_{PC} \sim P^\gamma$ with $\gamma$ = 3.35.

Super-linear photoresponse ($\gamma > 1$) reflects the exponential sensitivity of our photocurrent measurement to the excited state temperature $T_e$. Immediately after photoexcitation, the photon energy ($E_{PH}$ = 1.03 eV) is divided between a photoexcited electron and hole (Fig. 1d, inset). Since the initial charge carrier kinetic energy ($K = E_{PH} / 2 \sim 0.515$ eV) is significantly smaller than the hBN energetic barrier ($\Delta_h \sim 1.3$ eV for holes, $\Delta_e \sim 4.5$ eV for electrons[22,26]), direct transit is impossible. Instead, fast Auger-like scattering upconverts charge carriers[17,23] with sufficient energy to overcome the barrier via internal thermionic emission (Fig. 1d, inset). Charge carriers within the exponentially decaying tail of the hot distribution overcome the hBN barrier giving rise to strong super-linear dependence on laser power[22,26]. Since the energy barrier $\Delta_h < \Delta_e$, interlayer transit of hot carriers is mediated predominantly by holes[22,26].

Taking advantage of this super-linearity, we can probe the hot excited state within the first picoseconds after photoexcitation, establishing the typical response time of our technique. As shown in Fig. 1e, we resolve the dynamic, two-pulse photocurrent by separating identical laser pulses using a tunable time delay (see supplemental methods). When two gating pulses shine on the device simultaneous (time delay $\Delta t = 0$ ps), the resultant photocurrent is ten times greater than when the pulses are delayed by several picoseconds. As we increased $\Delta t$, $I_{PC}$ decreases exponentially with an ultrafast characteristic decay time $\tau = 1.31$ ps when a small interlayer voltage has been applied (red dashed line, Fig. 1f).

Having established the typical response characteristics, we next studied the interlayer photoresponse in the vicinity of the Dirac point voltage $V_D$ at room temperature (Fig. 2). As shown in Fig. 2a, the $I_{PC}$ vs. $V_B$ characteristics exhibit a sharp kink near $V_B = 0$ V. Comparing $I_{PC}$ to the device resistance, we find that this sharp photocurrent increase coincides directly with the charge neutrality (Dirac) point $V_D$ of the top graphene layer (Fig. 2b). The photocurrent remains super-linear with power at all voltages $V_B$ (Fig. 2c), exhibiting a voltage-tunable super-linearity. As a function of $V_B$, $\gamma$ increases gradually as we tune closer to $V_D$, abruptly collapsing in the same narrow voltage range over which the photoconductance $dI_{PC}/dV_B$ is sharply peaked (Fig. 2d).

While the super-linearity and interlayer photoconductance change dramatically at the Dirac point, the characteristic time dynamics also depend sensitively on $V_B$. Shown in Fig. 2e, the two-pulse inverse decay time $1/\tau$ scales approximately linearly with $V_B$. At the Dirac point, ultrafast interlayer charge transit from one graphene layer to the other sets the limiting response time, $\tau_D = 1.6$ ps. This response time bounds the time window accessed by our measurement (from ~ 180 fs to 1.6 ps) when the entire graphene layer is tuned to charge neutrality.

To infer $T_e$ from the voltage-dependent photoresponse, we model the photo-thermionic current that arises due to counter-propagating hot carriers between the top and bottom graphene layers (see supplemental methods). When $V_B$ is applied, an excess of hot carriers is driven in one direction by the interlayer electric field $\vec{\mathcal{E}}$ (schematic Fig. 2f). The density of hot carriers is uniquely determined by two parameters in each layer, $T_e$ and the chemical potential $\mu$. Since $\mu$ is determined by the quantum and geometric capacitances of the device, $\mu$ as a function of $V_B$ can be calculated for each layer. With $\mu$ determined at every $V_B$ value, and given the initial doping in graphene is small, we numerically determine the electronic temperature required to produce the observed photocurrent profile as a function of $V_B$. We thus extract the approximate temperature profile $T_e$ vs. $V_B$, shown in Fig. 2g, from the photocurrent characteristics (Fig. 2a).

The electronic temperature profile (Fig. 2g) provides a snapshot of the hot carrier regime. Fast carrier-carrier scattering dominates over electron-phonon scattering - particularly within the fast response time window of our device - and a greater fraction of the initial photon energy is captured by the ambient electronic system[20]. The sharply peaked electronic temperature profile is a distinct signature of this hot carrier transport regime[27]. As the device is tuned closer to $V_D$, the Dirac excited state reaches very high electronic temperature $T_e > 2000$ K.

This signature of strongly interacting hot carriers at room temperature offers an opportune benchmark against which we can compare the properties of the Dirac excited state at low temperatures. At each sample temperature $T$ we varied a small intralayer voltage $V_T$ between the two contacts of the top graphene layer $G_T$ while also applying $V_B$. From the photocurrent vs. $V_B$ and $V_T$ maps, we removed contributions to $I_{PC}$ that arise from ordinary changes of the in-plane conductivity (see supplemental methods). The remaining photocurrent occurs only across the interlayer barrier yet is sensitive to *in-plane* cooling processes in the top graphene layer.

In Figure 3, we compare the device photoresponse at $T = 143$ K to that at $T = 50$ K, where we observed markedly different behavior. At $T = 143$ K, the photoconductance vs. $V_B$ characteristics exhibit a single peak that changes *weakly* with $V_T$ (Fig. 3a). In contrast, at $T = 50$ K, the photoconductance-voltage characteristics evolve into a prominent multipeaked structure with deep valleys at the Dirac point (Fig. 3b). Within a narrow region around the Dirac point, the suppression of the photoconductance at $V_T = 0$ V (Fig. 3b, bottom) transitions into robust negative differential photoconductance as $V_T$ increases.

Suppression of the Dirac photoconductance provides a direct gauge of quenching in the hot excited state. At $V_B = V_D$, the thermionic transit of charge carriers is suppressed. As $V_B$ is tuned away from the Dirac point, the charge density and $\vec{\mathcal{E}}$ increase, resulting in positive photoconductance peaks just outside the Dirac region (Fig. 3d). Eventually, as $V_B$ is tuned further away from $V_D$, photoconductance decreases as heating the e-h population becomes less efficient at high ambient charge carrier density.

By examining the electronic temperature profile at $V_T = 0$ V (Fig. 4a), we determine the intrinsic quenching rate. When compared to the room temperature data (Fig. 2g), the $T_e$ vs. $V_B$ profile at $T = 50$ K exhibits a highly suppressed electronic temperature in the immediate vicinity of the Dirac point. By estimating the expected peak-to-valley electronic temperature change ($\Delta T_e \sim 285$ K) that occurs within $\tau_D = 1.6$ ps in Figure 4c, we infer a *lower* bound to the intrinsic cooling rate of 177 K/ps at the highest laser excitation power.

A subtle, but important, property of the quenching process is revealed in the detailed $T$ dependence of the Dirac photoconductance: Strong suppression at $T = 50$ K coincides directly with sensitive dependence on in-plane electrical bias. To see this, we tracked the photoconductance minima at the Dirac point as we varied $T$. As the sample temperature increased (Fig. 4b), the Dirac photoconductance at $V_T = 25$ mV dips to a sharp minimum at $T = 50$ K, reaching clear negative differential photoconductance. Above this temperature, the Dirac photoconductance rebounds sharply as the sample temperature approaches $T = 300$ K. As shown in Fig. 4c, the $V_T$-dependence at intermediate temperatures is significantly stronger than at high $T$. At $T = 50$ K, $V_T$ strongly enhances the cooling process, yet quenching of the Dirac excited state persists even in the absence of an in-plane electrical bias ($V_T = 0$ V).

Dirac excited state quenching observed here is highly unexpected, yet our measurements identify key features of this emergent cooling regime. At the critical sample temperature of $T = 50$ K, the electronic temperature of the Dirac excited state reaches $T_e > 10^3$ K but is rapidly quenched near the Dirac point (Fig. 4a). Before hot carriers can transit across the boron nitride interface, *in-plane* heat flow cools the excited state with remarkable efficiency. Quenching is directly controlled by the in-plane voltage (Fig. 4b). As the in-plane voltage increases, ultra-efficient cooling in the top layer $G_T$ leads to an interlayer backflow - and thus negative differential photoconductance (shaded regions Fig. 4) - of hot charge carriers from the bottom layer $G_B$. While quenching is most effective at intermediate sample temperatures, no quenching is observed at high sample temperatures.

It is unclear why quenching of the extremely high electronic temperature is so sensitive to sample temperature and in-plane voltage (see supplemental Figure S4). The striking temperature dependence indicates anomalously strong electron-vibrational coupling at short time scales, yet ordinary phonon cooling does not account for our observations. Optical phonon emission is known to contribute to hot carrier relaxation at short timescales[28,29], yet such high-energy emission processes (~180 meV per emitted phonon) are expected to be insensitive to sample temperature. In addition, the $T_e$ vs. $V_B$ snapshot taken within the early lifetime of the Dirac excited state (Fig. 4c) precludes the role of ordinary low-energy acoustic phonon scattering[13]; this energy relaxation process requires longer timescales than the picosecond response time of our technique. Moreover, a crossover between acoustic phonon scattering and the super-collision mechanism[14] is expected to lead to an *increase* of the electronic temperature at intermediate $T$ [15]. This is the opposite of the quenching process observed here.

Although voltage-sensitivity suggests that charge transport plays a role in the quenching process, the rapid energy relaxation observed here is *not consistent* with previous observations attributed to low-energy transport pathways. As a key comparison, enhanced thermal conductivity - ascribed to electrons and holes that move in the same direction - results from an energy current that efficiently reduces the electronic temperature, yet a total charge current that is precisely zero[6,8]. Enhanced $k_e$, however, is not expected to occur at elevated $T_e$, thus indicating energy relaxation processes beyond experimentally established low-temperature mechanisms. Theoretically, it is known that a number of mechanisms may contribute to rapid cooling and enhanced $k_e$; Examples include ballistic energy waves[30], hydrodynamic energy waves[4], Joule-Thomson processes[31], and phonon wind effects[32–34]. Future experiments will be required to elucidate these and other potential sources of the unexpected quenching process.

With high sensitivity to small changes in the electronic temperature - estimated here to exceed $dT_e/T_e \sim 4.75*10^{-2}$ at 2000 K (see supplemental methods) - and the ability to stack-engineer the interlayer energy barrier, our technique can be employed to study an increasing number of nanomaterials manifesting electron-hole liquids, solids, and plasmas. Indeed, interlayer optoelectronic thermometry overcomes a fundamental hurdle to probing transport through collective phases in ultrathin materials: Here, generating photocurrents is achieved without the implementation of lateral p-n junctions, whose ambient charge density masks intrinsic energy and charge transport. Beyond Dirac excited state quenching, other quantum critical phenomena in ultrathin van der Waals heterostructures may be explored, as one can optically probe $T_e$ within an embedded, intrinsic transport layer.

**Acknowledgements:**
We would like to acknowledge important discussions regarding both experimental and theoretical aspects of this work with Vivek Aji, Cory R. Dean, Leonid Levitov, Justin C. W. Song, and Xiaoyang Zhu

**Funding:**
This work was supported by the Army Research Office Electronics Division Award no. W911NF2110260 (N.M.G., J.C.W.), the Presidential Early Career Award for Scientists and Engineers (PECASE) through the Air Force Office of Scientific Research (award no. FA9550-20-1-0097; N.M.G. and T.B.A.), through support from the National Science Foundation Division of Materials Research CAREER Award (no. 1651247; N.M.G., T.B.A. and J.C.W.), and through the United States Department of the Navy Historically Black Colleges, Universities and Minority Serving Institutions (HBCU/MI) award no. N00014-19-1-2574 (N.M.G. and T.B.A.).

**Author contributions:**
J.C.W. contributed to the conceptualization of the optoelectronic thermometry technique, data analysis methodology, photocurrent measurements and Investigation, and writing, review, and editing of original draft.

T.B.A contributed to the conceptualization of the optoelectronic thermometry technique, data analysis methodology, photocurrent measurements and Investigation, and the review and editing of original draft.

N.M.G. contributed to the conceptualization of the optoelectronic thermometry technique, data analysis methodology, photocurrent measurements and Investigation, data visualization, funding acquisition, project administration and supervision, and writing, review, and editing of the original draft.

**Data and materials availability:**
All data, code, and materials used in the analysis can be made available upon reasonable request.

**References:**
[1] L. Fritz, J. Schmalian, et al., Quantum critical transport in clean graphene. *Phys. Rev. B* **78**, 085416 (2008).
[2] Daniel E. Sheehy and Jörg Schmalian *Phys. Rev. Lett.* 99, 226803 (2007).
[3] M. Müller, J. Schmalian, L. Fritz, Graphene: A Nearly Perfect Fluid. *Phys. Rev. Lett.* 103, 025301 (2009).
[4] W. Zhao, S. Wang, et al., Observation of hydrodynamic plasmons and energy waves in graphene. *Nature* **614**, 688-693 (2023)
[5] A. Lucas, J. Crossno, et al., Transport in inhomogeneous quantum critical fluids and in the Dirac fluid in graphene. *Phys. Rev. B* **93**, 075426 (2016).
[6] P. Gallagher, C. S. Yang, et al., Quantum-critical conductivity of the Dirac fluid in graphene. *Science* **364**, 158-162 (2019).
[7] A. I. Berdyugin, N. Xin, et al., Out-of-equilibrium criticalities in graphene superlattices. *Science* **375**, 430-433 (2022).
[8] J. Crossing, J. K. Shi, et al., Observation of the Dirac fluid and the breakdown of the Wiedemann-Franz law in graphene. *Science* **351**, 1058-1061 (2016).
[9] D. A. Bandurin, I. Torre, et al., Negative local resistance caused by viscous electron backflow in graphene. *Science*, **351**, 1055-1058 (2016).
[10] A. Lucas, K. C. Fong, Hydrodynamics of electrons in graphene. *J. Phys.: Condens. Matter* **30** 053001(2018).
[11] A. Block, A. Principi, N.C.H. Hesp, et al., Observation of giant and tunable thermal diffusivity of a Dirac fluid at room temperature. *Nat. Nanotechnol.* **16**, 1195–1200 (2021).
[12] N. Xin, J. Lourembam, P. Kumaravadivel, et al., Giant magnetoresistance of Dirac plasma in high-mobility graphene. *Nature* **616**, 270–274 (2023).
[13] J.C. W. Song, M. Y. Reizer, L. S. Levitov, Disorder-Assisted Electron-Phonon Scattering and Cooling Pathways in Graphene. Phys. Rev. Lett. **109**, 106602 (2012).
[14] M. Graham, S. F. Shi, D. Ralph, et al., Photocurrent measurements of supercollision cooling in graphene. *Nature Phys* **9**, 103–108 (2013).
[15] Q. Ma, N. M. Gabor, et al., Competing Channels for Hot-Electron Cooling in Graphene. *Phys. Rev. Lett.* **112**, 247401 (2014).
[16] Y. Lin, Q. Ma, et al., Asymmetric hot-carrier thermalization and broadband photoresponse in graphene-2D semiconductor lateral heterojunctions. *Science Adv.* **5**, DOI: 10.1126/sciadv.aav1493 (2019).
[17] I. Gierz, F. Calegari, et al., Tracking Primary Thermalization Events in Graphene with Photoemission at Extreme Time Scales. *Phys. Rev. Lett.* **115**, 086803 (2015).
[18] B. A. Ruzicka, S. Wang, et al., Hot carrier diffusion in graphene. *Phys. Rev. B* **82**, 195414 (2010).
[19] C. H. Liu, N. M. Dissanayake, et al., Evidence for Extraction of Photoexcited Hot Carriers from Graphene. *ACS Nano* **6,** 7172–7176 (2012).
[20] I. Gierz, J. Petersen, M. Mitrano, et al., Snapshots of non-equilibrium Dirac carrier distributions in graphene. *Nature Mater* **12**, 1119–1124 (2013).
[21] J. F. Rodriguez-Nieva, M. S. Dresselhaus, J. C. W. Song, Enhanced Thermionic-Dominated Photoresponse in Graphene Schottky Junctions *Nano Lett.* **16**, 6036–6041 (2016).
[22] J. F. Rodriguez-Nieva, M. S. Dresselhaus, L. S. Levitov, Thermionic Emission and Negative dI/dV in Photoactive Graphene Heterostructures. *Nano Lett.* 2015, 15, 1451–1456 (2015).

[23] D. Brida, A. Tomadin, C. Manzoni, et al., Ultrafast collinear scattering and carrier multiplication in graphene. *Nat Commun* **4**, 1987 (2013).
[24] L. Britnell, R. V. Gorbachev, et al., Electron Tunneling through Ultrathin Boron Nitride Crystalline Barriers. *Nano Lett.* 12, 1707−1710 (2012).
[25] A. Kuzmina, M. Parzefall, et al., Resonant Light Emission from Graphene/Hexagonal Boron Nitride/Graphene Tunnel Junctions. *Nano Lett.* **21**, 8332–8339 (2021).
[26] Q. Ma, T. Andersen, N. Nair, et al., Tuning ultrafast electron thermalization pathways in a van der Waals heterostructure. *Nature Phys* **12**, 455–459 (2016).
[27] N. M. Gabor, J. C. W. Song, et al. Hot Carrier–Assisted Intrinsic Photoresponse in Graphene. *Science* **334**, 6056 648-652 (2011).
[28] K. J. Tielrooij, N.C.H. Hesp, A. Principi, et al. Out-of-plane heat transfer in van der Waals stacks through electron–hyperbolic phonon coupling. *Nature Nanotech* **13**, 41–46 (2018).
[29] E. A. A. Pogna, X. Jia, et al. Hot-Carrier Cooling in High-Quality Graphene Is Intrinsically Limited by Optical Phonons. *ACS Nano* **15**, 11285–11295 (2021).
[30] T. V. Phan, J. C. W. Song, L. S. Levitov, Ballistic Heat Transfer and Energy Waves in an Electron System. arXiv:1306.4972v1 (2013) .
[31] K. Zarembo, Joule—Thomson Cooling in Graphene. *Jetp Lett.* **111**, 157–161 (2020).
[32] F. M. Steranka, J. P. Wolfe, Phonon-Wind-Driven Electron-Hole Plasma in Si. *Phys. Rev. Lett.* **53**, 2181 (1984).
[33] M. M. Glazov, Phonon wind and drag of excitons in monolayer semiconductors. *Phys. Rev. B* **100**, 045426 (2019).
[34] K. S. Bhargavi, S. S. Kubakaddi, Phonon-drag thermopower in an armchair graphene nanoribbon. *J. Phys.: Condens. Matter* **23**, 275303 (2011).
[35] N. Kharche, S. K. Nayak, Quasiparticle band gap engineering of graphene and graphene on hexagonal boron nitride substrate. Nano Lett. **11,** 5274_5278 (2011).

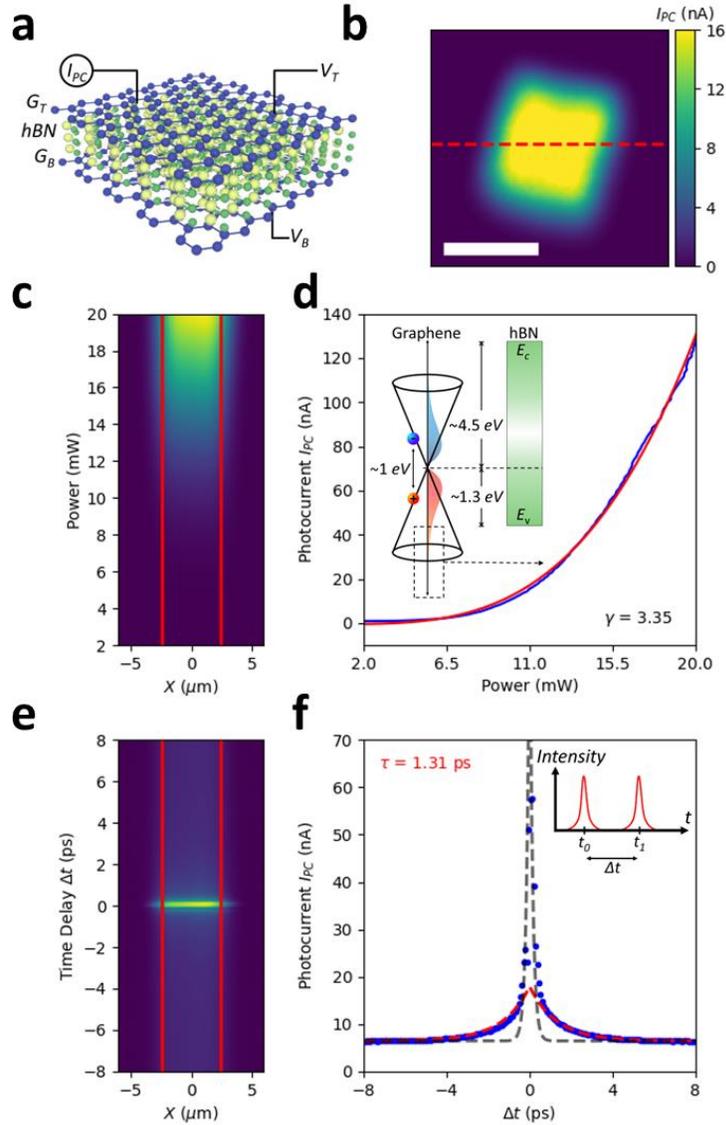

**Fig. 1 Photoresponse of G/hBN/G for interlayer optoelectronic thermometry. (a)** Schematic of encapsulated G/hBN/G devices. Graphene top layer ($G_T$) and graphene bottom layer ($G_B$) are separated by a thin h-BN layer. The voltage $V_B$ is applied to the bottom layer, top layer voltage $V_T$ = 0 V for data in Figures 1 and 2. **(b)** Spatial photocurrent map of the G/hBN/G overlap region, scale bar 10 um, wavelength $l$ = 1200 nm. **(c)** Interlayer photocurrent $I_{PC}$ vs. laser power taken across the red line shown in (b). **(d)** $I_{PC}$ vs. laser power (blue data), spatially averaged between the red lines in (c). The power law fit $I \sim P^\gamma$ (shown in red) shows a non-linearity factor of $\gamma$ = 3.35. Inset, schematic of hot carriers (holes) originating in graphene transiting the interlayer hBN energy barrier (dashed rectangle). **(e)** Two-pulse photocurrent vs. time delay $Dt$ across the red line shown in (b). **(f)** Two-pulse photocurrent (spatially averaged within the red lines in (e)) as a function of $Dt$. Characteristic decay time constant of 0.135 ps (pulse limited time constant) and 1.31 ps extracted from exponential fits (dashed lines).

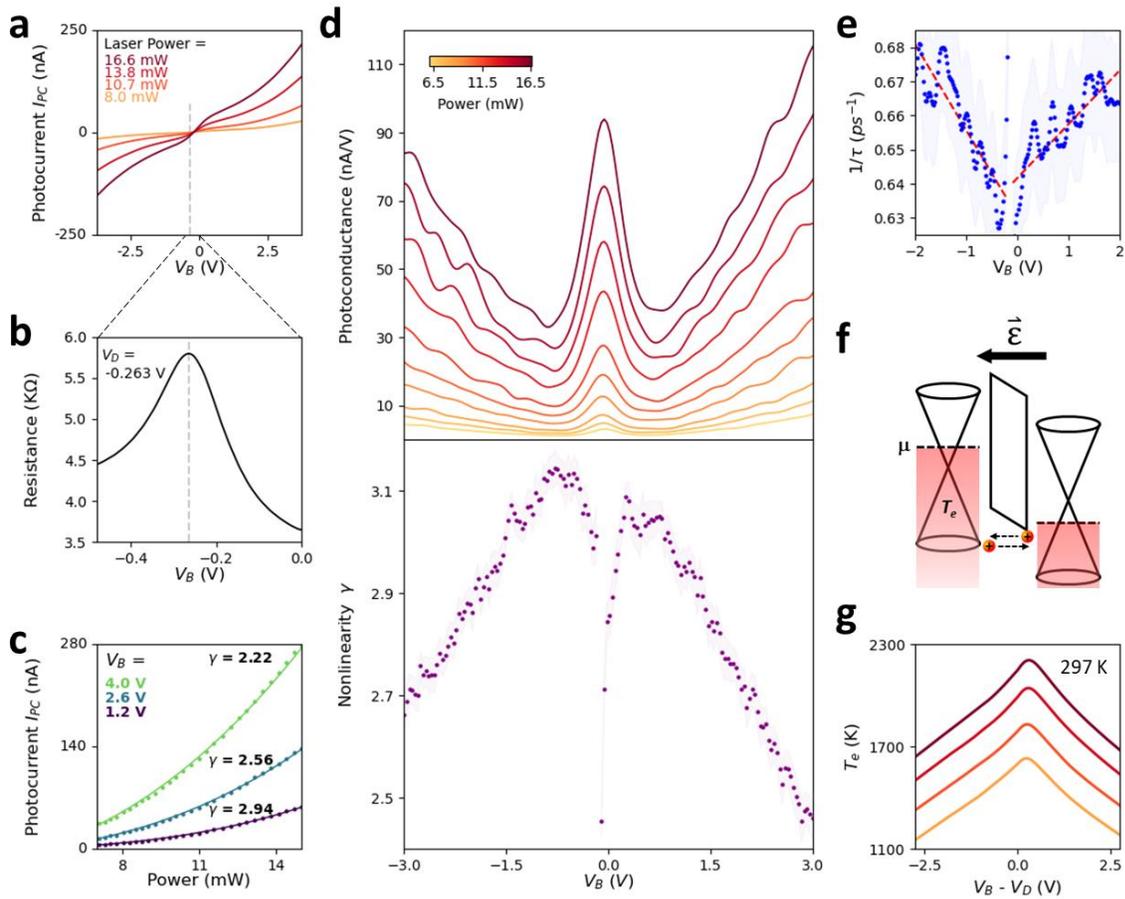

**Fig. 2 Interlayer optoelectronic thermometry of the hot Dirac excited state at room temperature. (a)** $I_{PC}$ vs. $V_B$ at various laser powers (labelled). The top graphene layer Dirac point is shown with dashed line. **(b)** Resistance (dark) of $G_T$ as a function of $V_B$, charge neutrality point at $V_B$ = -0.263V. **(c)** $I_{PC}$ vs. laser power at various $V_B$ values (labelled). The photocurrent remains super linear, exponents labelled. **(d)** Photoconductance and nonlinearity γ as a function of $V_B$. **(e)** Inverse decay time extracted from 2-pulse photocurrent measurements as a function of $V_B$. **(f)** Schematic of the interlayer photothermionic effect. **(g)** Extracted electronic temperature based on the photocurrent data in panel (a).

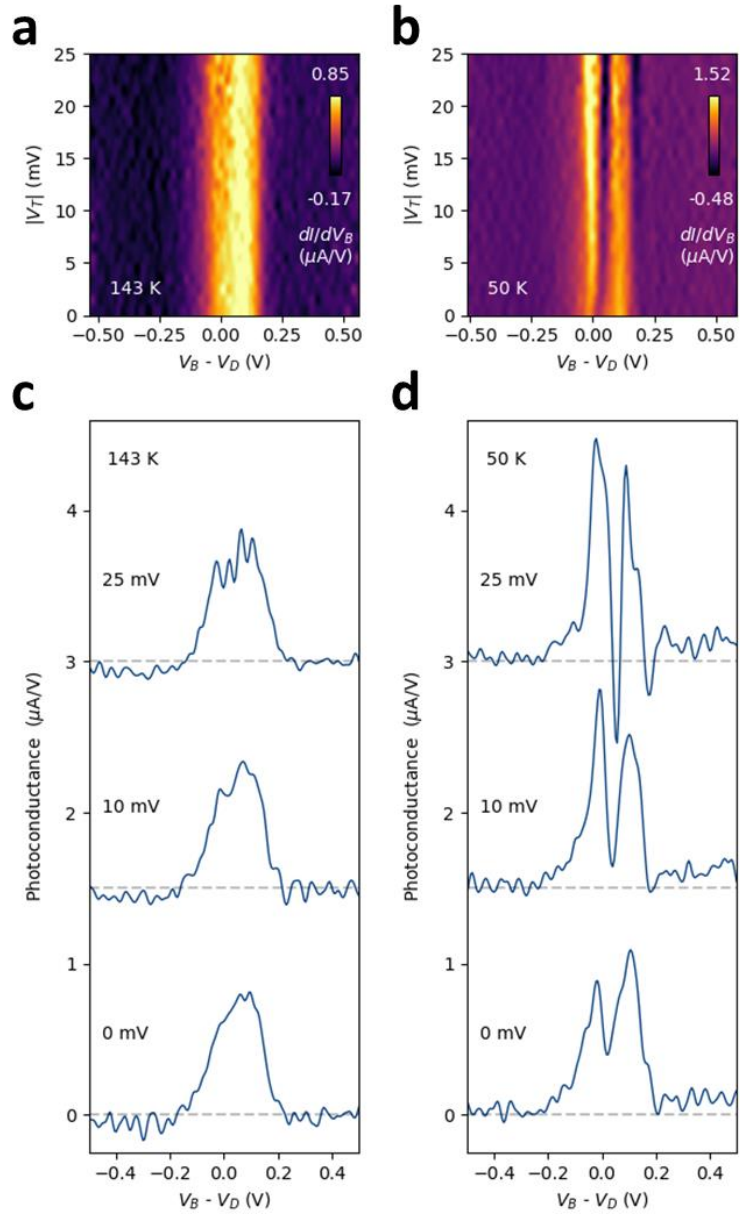

**Fig. 3 Dirac point photoconductance at high and low sample temperatures.** (**a**) Interlayer photoconductance $dI_{PC}/dV_B$ vs. $V_T$ and $V_B$ ($V_B$ has been shifted by the Dirac point voltage $V_D$) at $T$ = 143 K (See supplemental discussion for process used to isolate interlayer photoconductance from intralayer conductance) (**b**) Interlayer photoconductance vs. $V_T$ and $V_B$ at $T$ = 50 K. (**c**) Photoconductance vs. $V_B$ for $T$ = 143 K at several intralayer voltages (labelled and offset, zero marked by dashed line). (**d**) Photoconductance vs. $V_B$ for $T$ = 50 K at several intralayer voltages.

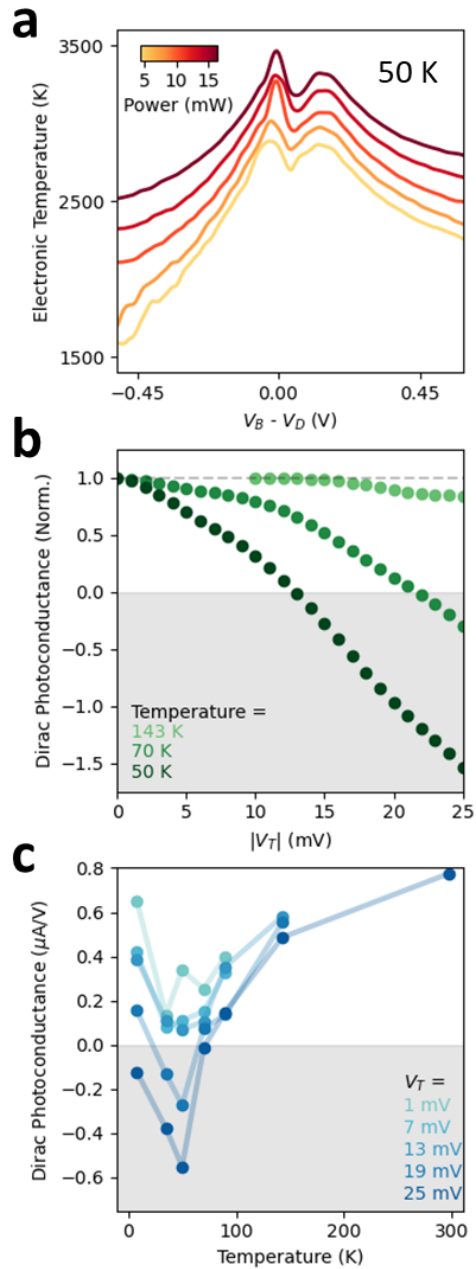

**Fig. 4 Suppression of the hot Dirac excited state photoconductance at intermediate sample temperatures.** (a) Extracted electronic temperature for photocurrent at $T = 50$ K and $V_T = 0$ mV, at several values of laser power. (b) Anomalous dependence of Dirac point photoconductance on interlayer voltage $V_T$ compared at several temperatures. (c) Dirac point photoconductance vs. $T$ at several values of interlayer voltage $V_T$.

## Supplementary Materials

### Interlayer Optoelectronic Thermometry

We build an electronic thermometer by using graphene and its charge carriers in the place of glass and an encased medium (e.g., liquid mercury) as in a conventional thermometer. In a conventional thermometer, the medium absorbs heat from its environment and expands which causes the level of the medium to increase in height. Thus, the temperature can be determined by height reached by the medium. We can apply a similar concept to charge carriers in graphene, in which the medium is now the charge carriers (Fig. S1). When the charge carriers in graphene absorb energy, they can thermalize and reach a quasi-equilibrium that is at a higher temperature that the lattice[17-20]. Within the thermalized distribution, many charge carriers will reach a higher energy than the initial excited population. By introducing an out-of-plane energy barrier we can filter out the high energy tail of this distribution[21-24] and extract them as a photocurrent. The magnitude of this photocurrent is directly related to the number of carriers with high enough energy to overcome the energy barrier which can then be related to the electronic temperature of this system.

### Generalized Chemical Potential Calculation

We determine the initial doping by considering the capacitive coupling between layers in our G-hBN-G heterostructure and by adapting a similar approach as reference[22]. As we apply an interlayer voltage to $G_B$, the quantum capacitance of graphene describes the change in the charge density (thus chemical potential), while the geometric capacitance describes the potential drop across the two graphene layers. The interlayer voltage is the sum of the change in chemical potentials and the potential drop across the layers.

A general relationship between the interlayer voltage, electric field between the graphene layers, and the chemical potential in each graphene layer can be written (Eq. S1), where $e$ is the electron charge, $V_B$ is the interlayer voltage, $\mu_T$ is the chemical potential of the top layer of graphene, $\mu_{T0}$ is the initial doping (chemical potential) of the top layer of graphene, $\mu_B$ is the chemical potential of the bottom layer of graphene, $\mu_{B0}$ is the initial doping of the bottom layer of graphene, $E$ is the electric field between the two graphene layers, and $d$ is the separation between the graphene layers (the interlayer hBN thickness). We can relate the chemical potentials and the electric field by the change in charge density such that we end up with a relationship between $V_B$ and either $\mu_T$ or $\mu_B$. To do this we must first determine the relationship between $\mu_T$ and $\mu_B$.

$$eV_B = \mu_T - \mu_{T0} + \mu_{B0} - \mu_B + eEd \quad (S1)$$

$$V_B = \mu_T - \mu_{T0} + \mu_{B0} - \mu_B + Ed \quad ; in\ units\ of\ eV \quad (S2)$$

When an interlayer voltage is applied, the change in charge density in one layer is equal and opposite that of the other layer (Eq. S3). Given the density of states of graphene, we can express the chemical potential in terms of the change in charge density (Eq. S4-S6). By combining equation 3 and equation 6 we can express the relationship between $\mu_T$ and $\mu_B$ (Eq. S7). We can then express $\mu_T$ as a function of $\mu_B$ and vice versa (Eq. S8 and Eq S9). The electric field, $E$, between the layers can now also be expressed in terms of the chemical potential if we consider the two graphene layers as a capacitor with a dielectric, hBN. We can now rewrite equation S2 in terms of either $\mu_T$ or $\mu_B$ (Eq. S10 and Eq. S11 respectively). Although there is still the sign of the opposing layer to consider in these two expressions, they are entirely determined by whether or not the quantity under the square root is positive, which only depends on the layer in question. We then can solve for the combination of $\mu_{T0}$ and $\mu_{B0}$ which simultaneously satisfies both the top layer and bottom layer Dirac point voltage, $V_{D,T} = V_B(\mu_T = 0)$, $-V_{D,B} = V_B(\mu_B = 0)$, which yields the initial doping of both layers.

$$\Delta n_T = -\Delta n_B \quad (S3)$$

$$\mu_i = -\frac{\hbar v_F}{e}\sqrt{\pi|n_0 + \Delta n_i|}sign(n_0 + \Delta n_i) \quad (S4)$$

$$sign(\mu_i)\mu_i^2 = -\frac{\hbar^2 v_F^2}{e^2}\pi(n_0 + \Delta n_i) \tag{S5}$$

$$sign(\mu_i)\mu_i^2 = sign(\mu_{i0})\mu_{i0}^2 - \frac{\hbar^2 v_F^2}{e^2}\pi(\Delta n_i) \tag{S6}$$

$$sign(\mu_T)\mu_T^2 - sign(\mu_{T0})\mu_{T0}^2 = -sign(\mu_B)\mu_B^2 + sign(\mu_{B0})\mu_{B0}^2 \tag{S7}$$

$$\mu_T = sign(\mu_T)\{sign(\mu_T)[-sign(\mu_B)\mu_B^2 \tag{S8}$$
$$+ sign(\mu_{T0})\mu_{T0}^2 + sign(\mu_{B0})\mu_{B0}^2]^{1/2}\}$$

$$\mu_B = sign(\mu_B)\{sign(\mu_B)[-sign(\mu_T)\mu_T^2 \tag{S9}$$
$$+ sign(\mu_{T0})\mu_{T0}^2 + sign(\mu_{B0})\mu_{B0}^2]^{1/2}\}$$

$$V_B(\mu_T) = \mu_T - \mu_{T0} + \mu_{B0} \tag{S10}$$
$$- sign(\mu_B)\{sign(\mu_B)[-sign(\mu_T)\mu_T^2$$
$$+ sign(\mu_{T0})\mu_{T0}^2 + sign(\mu_{B0})\mu_{B0}^2]\}^{1/2}$$
$$- \frac{de^3}{\pi\hbar^2 v_F^2}[sign(\mu_T)\mu_T^2 - sign(\mu_{T0})\mu_{T0}^2]$$

$$V_B(\mu_B) = -\mu_B - \mu_{T0} + \mu_{B0} \tag{S11}$$
$$+ sign(\mu_T)\{sign(\mu_T)[-sign(\mu_B)\mu_B^2$$
$$+ sign(\mu_{T0})\mu_{T0}^2 + sign(\mu_{B0})\mu_{B0}^2]\}^{1/2}$$
$$+ \frac{de^3}{\pi\hbar^2 v_F^2}[-sign(\mu_B)\mu_B^2 + sign(\mu_{B0})\mu_{B0}^2]$$

When calculating the initial doping, we first determine the Dirac point voltage, $V_D$, for each layer of graphene. This is done by measuring the in-plane resistance across one of the graphene layers by applying a source voltage to one of the contacts and measuring the dark current on the opposing contact while using the other layer as a gate. The maximum of the resistance peak shows the Dirac Point voltage, $V_B(\mu_i = 0) = V_{D,i}$, for each layer of graphene. All of our devices show a resistance peak close to 0 V of interlayer voltage which suggests they are close to being charge neutral. This can be confirmed in the following calculation for the initial doping.

Using equation S10 and S11, we first select an arbitrary test $\mu_{T0}$ and solve for $V_{D,T}$ or $V_{D,B}$ over a series of test $\mu_{B0}$ values. By interpolating $\mu_{B0}$ as a function of $V_D$ we can determine the value of $\mu_{B0}$ with the chosen $\mu_{T0}$ which yields the correct Dirac point voltage for each equation. This is repeated over a series of $\mu_{T0}$. The result are two sets of $\mu_{T0}$ and $\mu_{B0}$ combinations in which one set satisfies $V_{D,T}$ and the other satisfies $V_{D,B}$. By plotting $\mu_{T0}$ as a function of $\mu_{B0}$ the difference between the two curves can be determine, thus the combination of $\mu_{T0}$ and $\mu_{B0}$ which has no difference can be found numerically and satisfies both $V_{D,T}$ and $V_{D,B}$ simultaneously. Table S1 shows the calculated initial doping. Device B are calculated using the Dirac point voltage obtained from the minimum dark current transconductance of both layers of graphene. For Device A, the bottom layer only has one working contact, therefore no dark current was measured for the bottom layer and we assume the Dirac point voltage is the same as the top layer. This is a reasonable assumption since both layers of a given device are made in a similar time frame and stored under the same conditions which yields similar doping as seen in Device B

**Photocurrent Modeling and $T_e$ Estimation**

The photocurrent depends on the chemical potential and the electronic temperature at a specific interlayer voltage (Fig. *S2*). The total photocurrent $I_{PC}$ is the sum of the counter propagating photocurrent originating from each layer of graphene $I_{PC} = I_B - I_T$ where $I_B$ and $I_T$ are the current from the bottom and top layer respectively. The magnitude of the photocurrent from each layer depends on the population of carriers with high enough energy to overcome the hBN barrier. We calculate this by integrating all carriers with energy larger than the barrier, $U_0 = 1.3$eV [26,35]. Since the photocurrent is expected to be dominated by the transport of holes, we shall integrate the population of holes up to the valence band energy, $-U_0$, from negative infinity (Eq. S12).

$$I_i = e \int_{-\infty}^{-U_0} \rho(\varepsilon) f(-\varepsilon + \mu_i) \, d\varepsilon \tag{S12}$$

$$I_i = C \int_{-\infty}^{0} |\varepsilon - U_0| f(-\varepsilon + U_0 + \mu_i) \, d\varepsilon \tag{S13}$$

Here, $e$ is the electron charge, $\rho(\varepsilon)$ is the density of states of graphene as a function of energy, $f$ is the Fermi Dirac Distribution, and $\mu_i$ is the respective chemical potential for the *i*th layer (i = Top or Bottom). We can perform a change of variable to change the integration limit from $-U_0$ to 0 such that the integral has known analytic solutions (Eq. S13). The full integral is expressed in equation 14. Here, $C$ captures all the constants from the density of states and the electron charge, $C = (2e/\pi\hbar^2 v_F^2)$, $k_B$ is the Boltzmann Constant, and $T_e$ is the electronic temperature. Solving both integrals, we get Equation S15, where $Li$ is a polylogarithmic function. We rewrite the total photocurrent in terms the two parameters, $T_{e,i}$ and $\mu_i$ (Eq. S16). Since the chemical potential can be determined at any interlayer voltage value (see "generalized chemical potential calculation" section), this photocurrent model allows us to estimate the electronic temperature from our photocurrent data.

$$I_i = C \int_{-\infty}^{0} \frac{-\varepsilon}{e^{\beta(U_0+\mu_i)} e^{-\beta\varepsilon}} + \frac{U_0}{e^{\beta(U_0+\mu_i)} e^{-\beta\varepsilon}} \, d\varepsilon \quad ; \beta = \frac{1}{k_B T_{e,i}} \tag{S14}$$

$$I_i = C \left[ \frac{1}{\beta^2} Li_2\left(e^{-\beta(U_0+\mu_i)}\right) + \frac{1}{\beta} \ln\left(e^{-\beta(U_0+\mu_i)} + 1\right) \right] \tag{S15}$$

$$I_{PC} = I_B(\mu_B T_{e,B}) - I_T(\mu_T T_{e,T}) \tag{S16}$$

The electronic Temperature profile is estimated by determining the temperature required to match the profile our photocurrent data at each data point. Since the initial doping of our devices are relatively small, we will consider the purely symmetric case in which the chemical potentials in each layer start off at 0 eV. In this case, as the interlayer voltage is varied, $\mu_B$ and $\mu_T$ change in opposite directions but remains equal in magnitude. This determines the relationships between the applied interlayer voltage, $V_B$, and the chemical potentials in each layer, which will be equal in magnitude but opposite in sign. This allows us to simplified Equation S16 to one general chemical potential $\mu(V_B)$. Since the electronic temperature is the result of the charge carriers retaining the energy from photoexcitation, the final temperature would depend only on the electronic heat capacity of graphene, in other words it would depend on the available density of states at a given chemical potential. This implies the final electronic temperature should be symmetric about the charge neutrality point in each layer, and in the case of a symmetrically varying chemical potential, the electronic temperature is expected to be the same in each layer. This gives us a form of the interlayer photocurrent which depends on a single $T_e(V_B)$ which can be numerically fitted to match the appropriate photocurrent at a specific $V_B$ (eq. S17).

$$I_{PC}(V_B) = I_B(\mu(V_B), T_e(V_B)) - I_T(-\mu(V_B), T_e(V_B)) \tag{S17}$$

With the photocurrent measured and the chemical potential determined by electrostatics, the electronic temperature can now be solved for numerically. At each interlayer voltage value, the chemical potential is first calculated, and a series of test values for $T_e$ is used to calculate a series of possible $I_{PC}$. By interpolating the test $I_{PC}$ as a function of test $T_e$, we can then solve for the $T_e$ which yields the measured photocurrent. Due to the inherent asymmetry of the measured photocurrent, we must make an approximation of the Dirac point location $V_D$ while maintaining the profile of the photocurrent. We first center the data to the Dirac point of the layer from which we are measuring the photocurrent, such that $V_B - V_D = 0\,V$ in our model corresponds to the Dirac point of the top layer of graphene. The photocurrent is then offset so that it is zero at $0\,V$, since the model necessitates 0 A of photocurrent when the system is charge neutral. This process maintains the profile and changes of the photocurrent and only makes an approximation of the Dirac point voltage in the data. From this basis we can extract the electronic temperature and the manner which it varies with $V_B$.

**Device Fabrication**

We fabricated an encapsulated, nearly intrinsic and almost electronically symmetric double graphene layer device for our experiment. We first create our electrical contacts on our substrate then later transfer our material on to them. This will minimize possible sources of residue and damage from the electron beam lithography steps. 4 pairs of Titanium/Gold contacts are prefabricated on to a silicon substrate by electron beam lithography and are arranged in a rectangular manner. The substrate is a P-doped silicon wafer which has a surface consisting of a ~300 nm thick $SiO_2$ thermal oxide layer. These contacts are between 30nm to 50nm in height.

To minimize the time our materials encounter atmosphere we have built a simple motorized machine for our mechanical exfoliation process to maximize the amount of individual usable graphene and hBN flakes in a fixed amount of time. And each layer is transferred as they become available to minimize exposure to atmosphere.

We first find and transferred the base layer of encapsulating hBN via a standard dry transfer method. The hBN is first exfoliated then picked up with a stamp. The stamp consists of a glass slide with a layer of PDMS layered on top, then a layer of PPC is spun coated and cured on top of the PDMS at 180 degrees Celsius. The material is picked up by putting the PPC in contact with the material and heating pass the glass transition temperature of PPC (~45°C) to allow the stamp to come into good contact with the material and lifted off after cooling back down to (30°C). The next layer of graphene is transferred by first picking up a layer of hBN then using the hBN to pick up the layer of graphene. For the device discussed here this layer of hBN is between 7 nm and 10 nm in thickness. The layers are then laid on top of the base hBN and contacting the designated pair of contacts. This step is repeated for the second layer of graphene which is layered on top in a cross geometry to contact the other two pairs of electrical contacts. The result is a device in which the graphene is in a cross geometry and the heterostructure region in the center is encapsulated.

**Two-Pulse Scanning Laser Microscope**

We generate hot electron-hole excited states in our devices using ultra short, infrared optical gating pulses (Fig S3). We utilize a MIRA 900 optical parametric oscillator tuned to the wavelength of 1200 nm to photoexcite our devices. This laser outputs femtosecond (180 fs) pulses at a repetition rate of 76 MHz. The output passes through a 50/50 beam splitter which separates the beam into a reference path and a delay path. The delay path passes through a delay stage which varies the path length thus creating a time delay $\Delta t$ between the pulses. The pulses are then recombined with another 50/50 beam splitter to reform a single beam. Both paths include a half waveplate such that the two paths will be cross polarized to prevent interference of the pulses when recombined. The combined beam intersects an optical chopper that allows us to record the photoresponse using a Lock-In amplifier to measure the photoresponse with higher sensitivity and also remove any non-photoinduced responses. Finally, the beam arrives at a 90/10 beam splitter in which 10 % of the beam passes straight through the beam splitters and is collected with a InGaAs photodetector. The 10% of light collected at the InGaAs detector is used as a reference for the total power output of the laser. The 90% of the beam is reflected and redirected to a set of scanning optics for our microscopy measurement.

The scanning optics utilize a rotating galvo mirror and two lenses to control the light's angle of incidence onto the final objective, in this case a GRIN lens, thus changing the position which the laser focus on the device. In principle, the galvo mirror is set such that the center of rotation is at the focal point. The collimated beam, when incident upon the first lens, will then travel along a horizontal path but converge, with the point of convergence set at the focal length of the second lens. When the converging beam incident upon the second lens, the beam is once again collimated but now redirected to travel towards the focal point of the second lens. Thus, depending on the angle of the rotating mirror, we vary the angle of incidence of a collimated beam onto the back of our objective/Gradient Index of Refraction (GRIN) lens. This allows us to spatially scan the diffraction limited beam spot of our laser.

**Isolating intralayer and interlayer photoresponse**

When probing the in-plane cooling dynamics we may apply an in-plane voltage ranging from -30mV to 30mV. The photocurrent is mapped out as a function of $V_T$ and $V_B$. Since this in-plane voltage can itself drive excess current due to the change in the in-plane conductivity originating from photoexcitation. To isolate the interlayer portion of this response, we sum the photocurrent data from the $V_T$ = -30 mV case with the $V_T$ = 30 mV case. This will remove the component of the current that is directly driven by the in-plane voltage.

**Temperature Sensitivity Estimation**

We expect the photocurrent to be a sensitive probe of the electronic temperature due to the combination of the nonlinear relationships between the photocurrent and the incident laser power, and the nonlinear relationship between the laser power and the electronic temperature. We can estimate our sensitivity to the electronic temperature by examining the relationship between the electronic temperature and excitation power, relating the excitation power to the measured photocurrent thus giving us the connection between our measured photocurrent and the electronic temperature.

Based on the extracted electronic temperature at different laser power we estimate the relationship between them is approximately $T_e \sim 0.694 P^{0.411}$, where $P$ is the excitation power (in mW). Having determined from our photocurrent data that $I_{PC} \sim P^3$ and the exponent is the nonlinearity factor of the photocurrent (in nA), we can determine that $T_e \sim 0.694 I_{PC}^{0.137}$ The measurable resolution of our electronic temperature $\Delta T_e = \Delta I_{PC}*(dT_e/dI_{PC}) = \Delta I_{PC}*(0.694 *0.137)*(I_{PC})^{(-0.863)}$ where $T_e$ is in units of $10^3$ K, $I_{PC}$ is in units of $10^{-9}$ A, and $\Delta I_{PC}$ is the resolution of our photocurrent measurement $10^{-11}$ A (in other words $10^{-2}$ nA) Given the typical photocurrent near the Dirac point is approximately $10^{-9}$ A, this means $\Delta T_e \sim$ 95 K at 1nA and 1.79 K at 100nA and a sensitivity $\Delta T_e/T_e \sim$ 4.75x $10^{-2}$ at 2000K.

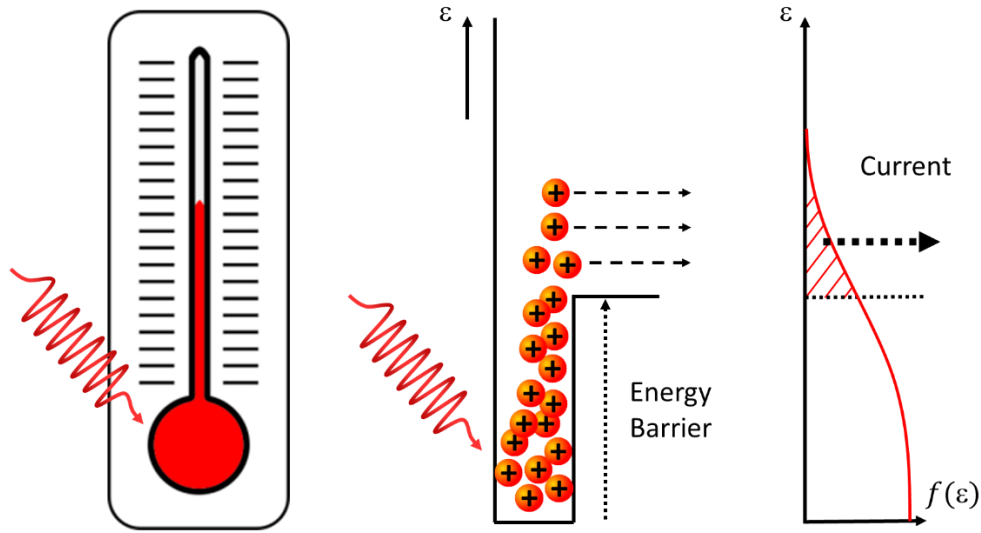

**Fig. S1.**

**Graphene Thermometer**. A conventional thermometer is represented as an analog to a graphene thermometer in which when a medium, alcohol/charge carriers, increases in elevation/energy level, when heated.

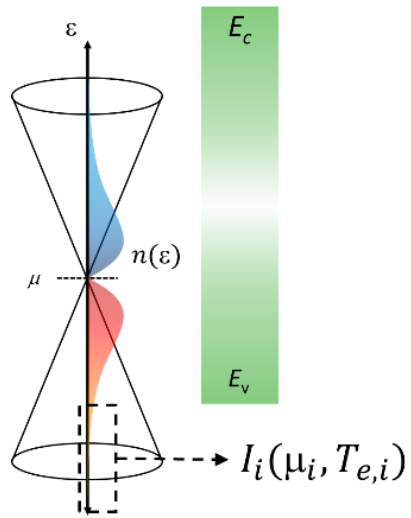

**Fig. S2.**

**Photocurrent Parameters.** The population of carriers above the barrier depends on the shape of the final thermalized carrier distribution which depends on the position of the chemical potential at the time of photoexcitation and the electronic temperature after thermalization.

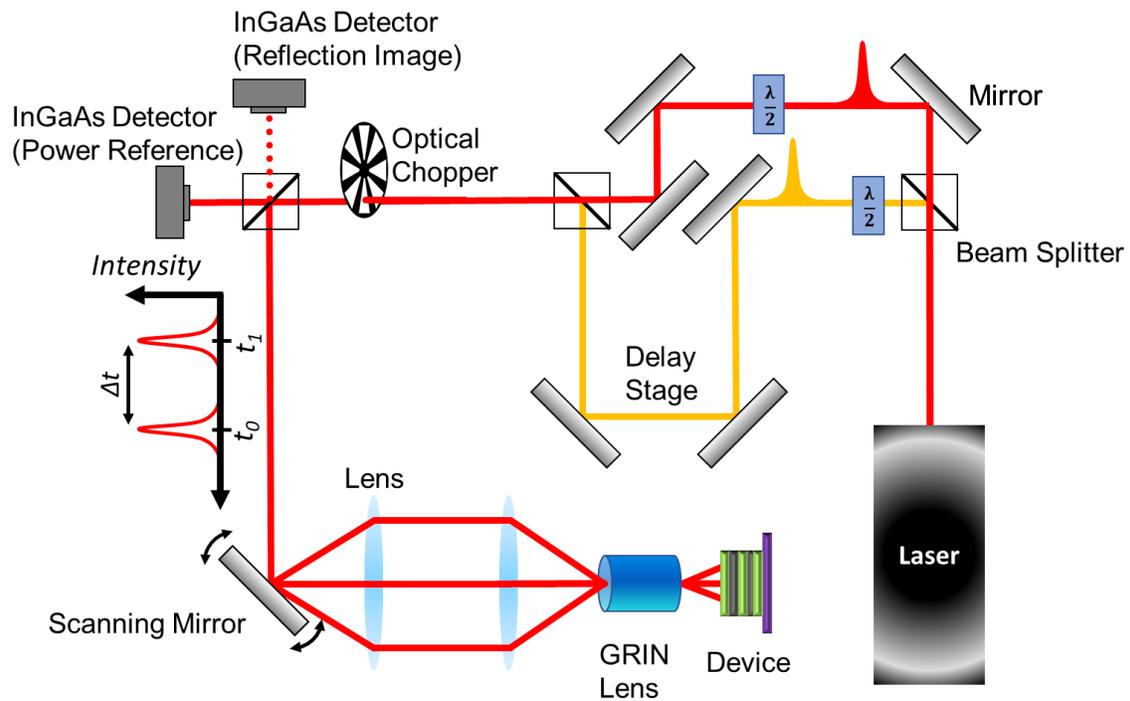

**Fig. S3.**
**Scanning Photocurrent Microscope**. An ultrafast pulse laser is used to optically excite our devices. The laser pulses are optically delayed on a controlled delay stage, and the laser's focal position can be varied via a set of scanning optics comprised of a rotating mirror, 2 lenses and a GRIN lens as the final focusing optic.

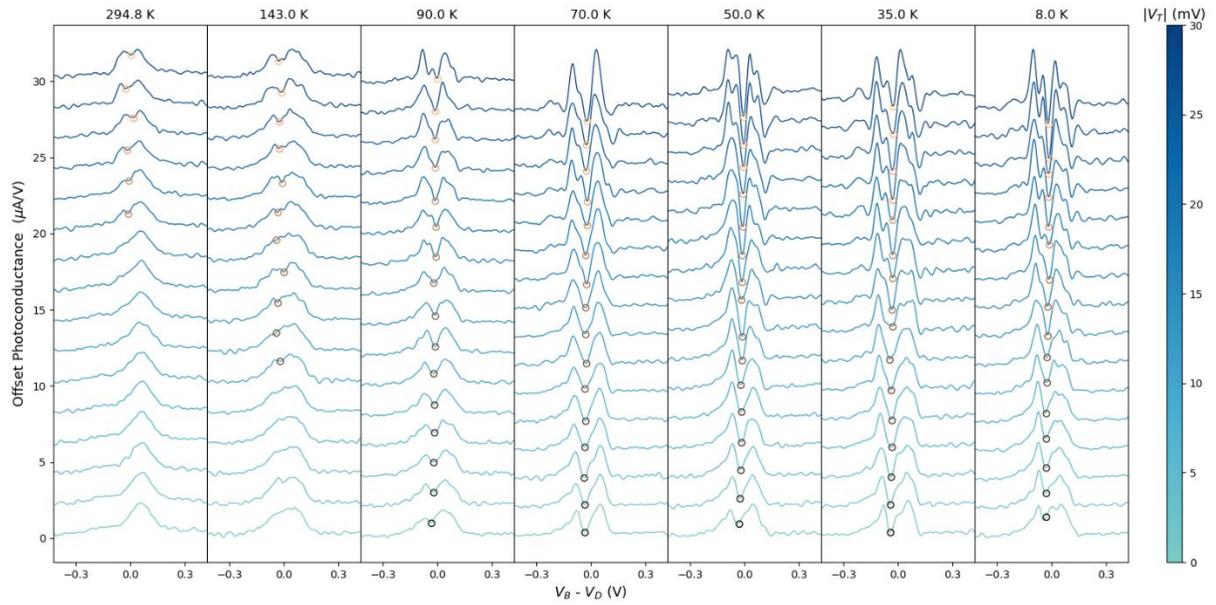

**Fig. S4.**
**Interlayer Photoconductance Suppression**. The photoconductance of the extracted interlayer photocurrent at various intralayer voltage (vertically offset) and lattice temperature is plotted as a function of interlayer voltage relative to the Dirac point voltage. The onset of the suppression of the photoconductance occurs at smaller values of $V_T$ at lower temperatures, and is the most significant around 50 K. The position of this suppression is marked with a circle.

| Device | $\mu_T$ (eV) | $\mu_B$ (eV) |
|--------|--------------|--------------|
| A      | 0.173        | 0.173        |
| B      | -0.096       | -0.053       |

**Table S1.**

**Calculated Initial Doping**. Calculated initial doping for devices used in the extraction of electronic temperature presented in this study. Device A is used in data presented in figure 1 and 2 of the main text, while device be is used in data presented in figure 3 and 4 which shows the quenching of the Dirac excited state.